%% file: main.tex
\documentclass[submission, Phys]{SciPost}
\usepackage[bitstream-charter]{mathdesign}
\usepackage[british]{babel}
\usepackage{braket}
\usepackage{mathtools}
\usepackage[T1]{fontenc} 
\usepackage[normalem]{ulem}
\usepackage{listings}
\usepackage{cleveref}
\usepackage{todonotes}
\usepackage{orcidlink}
\usepackage{comment}
\usepackage{subcaption}
\newcommand{\ve}[1]{\boldsymbol{#1}}
\usepackage{hyperref}
\begin{document}

\begin{center}{\Large \textbf{
Analytic continuation of Green's functions with a neural network
}}\end{center}

\begin{center}
Fakher Assaad \orcidlink{0000-0002-3302-9243},
Johanna Erdmenger 
\orcidlink{0000-0003-4776-4326},
Anika Götz,
René Meyer \orcidlink{0000-0002-6607-8199},
Martin Rackl\textsuperscript{$\star$} and \\
Yanick Thurn\textsuperscript{$\dagger$}
\orcidlink{0000-0001-8242-3362}
\end{center}

\begin{center}
 Institute for Theoretical Physics and Astrophysics and \\W\"urzburg-Dresden Cluster of Excellence ct.qmat\\ Julius-Maximilians-Universit\"at W\"urzburg\\ 97074 W\"urzburg, Germany
\\
$\star$ \href{mailto:email1}{\small martin.rackl@uni-wuerzburg.de}\,,\quad
$\dagger$ \href{mailto:email2}{\small yanick.thurn@uni-wuerzburg.de}
\end{center}

\begin{center}
\today
\end{center}


\section*{Abstract}
{
An important problem in many-body physics is to reconstruct the spectral density from the imaginary-time domain  Green's function. 
Typically, the imaginary-time Green's function is generated by Monte Carlo methods. As the one-point fermionic kernel diverges exponentially for  large frequencies, numerical noise generically causes instabilities. 
We use a convolutional neural network to obtain the spectral density for a given imaginary time Green's function. 
The network is trained by data which we generate using random Gaussians. 
We improve the training data set available by including  collision centers for the Gaussians rather than employing uniformly distributed Gaussians. 
Our network is constructed in such a way that its output fulfills  positive semidefiniteness. 
We compare the results of our network with results of the Maximum Entropy method (MaxEnt), a standard method for the same reconstruction problem for the spectral density. 
This comparison is performed for three different cases, namely our Gaussian based test data as well as two physical models,  the 1d Hubbard model showing spin-charge separation, and the two-dimensional SSH model in the self-consistent Born approximation. 
We find that the network outperforms   MaxEnt when presented data close to the training set.  
For the physical models considered, MaxEnt recognizes physical features more precisely as compared to our network prediction.  
While it is hard to improve MaxEnt, the quality of the network depends on the training data set which can be systematically enhanced and improved.}

\vspace{10pt}
\noindent\rule{\textwidth}{1pt}
\tableofcontents\thispagestyle{fancy}
\noindent\rule{\textwidth}{1pt}
\vspace{10pt}

\input{chapters/introduction}

\input{chapters/methods}

\input{chapters/performance}
\input{chapters/discussion}

\section*{Acknowledgements}
We thank Robert Helling for useful discussions. 

\paragraph{Author contributions}
The core methodology and neural network framework were developed and implemented by Martin Rackl and Yanick Thurn. Data for testing and Maximum Entropy analysis were contributed by Fakher Assaad and Anika Götz. Johanna Erdmenger and René Meyer contributed through scientific consultation and conceptual guidance.
All authors contributed to the writing of the manuscript.

\paragraph{Funding information}
The authors acknowledge financial support by the Deutsche Forschungsgemeinschaft (DFG, German Research Foundation) through the Würzburg-Dresden Cluster of Excellence ctd.qmat – Complexity, Topology and Dynamics in Quantum Matter (EXC 2147, project-id 390858490), and through the Collaborative Research 
centre “ToCoTronics”, Project-ID 258499086-SFB 1170. We also acknowledge financial support by the DFG through the German-Israeli ProjectCooperation (DIP) grant ‘Holography and the Swampland’.



\bibliography{SciPost_Example_BiBTeX_File.bib}

\nolinenumbers

\end{document}

%% file: chapters/introduction.tex
\section{Introduction}
Some problems in theoretical physics are easy to solve in one direction, but the inverse problem is hard, in particular it is prone to instabilities. 
One example is the relation between imaginary time domain Green's function and real frequency Green's function. 
Solving for the Green's function in the real frequency domain from euclidean domain lattice data is however one of the most important problems both in the fields of high energy and condensed matter physics, as many strongly interacting problems in these fields can only be solved by euclidean lattice simulations. 
Solving  this analytic continuation problem is for example  important for the determination of the transport coefficients, such as the shear viscosity \cite{PSS,KSS,teaney2010viscous}, of the quark gluon plasma, or for understanding spectroscopic experiments  such as neutron scattering, angle-resolved photoemission, or  scanning tunneling microscopy   in the realm of the solid state \cite{Negele,Sachdev:2010uz}. 
This analytic continuation is a so-called ill-posed problem, as it is resistant to analytical solution and simple numerical approaches, and lacks essential properties such as uniqueness or stability of solutions \cite{Koji}.
The crucial aspect of this problem is the instability of calculating the real frequency domain spectral density function $A(\omega) = \frac{1}{\pi}\text{Im}(G^R(\omega))$ from the imaginary time domain Green's function $\mathcal{G}(\tau)$ via
\begin{equation}
    \label{eq:ill-posed-problem}
    \mathcal{G}(\tau) = - \int d\omega \frac{e^{-\omega \tau}}{1 + e^{-\frac{\hbar \omega}{k_B T}}}A(\omega) = - \int d\omega\, \mathcal{K}(\tau, \omega) A(\omega),
\end{equation}
with $\mathcal{K}(\tau, \omega)$ being the integral kernel for a fermionic particle  and $ {\cal G}(\tau) = - \langle T \hat{c}_{x}(\tau) \hat{c}^{\dagger}_{x} \rangle $ is  the  single particle  Green function in euclidean space-time  for  a  fermionic excitation with quantum numbers  $x$.\footnote{For Bose  statistics the distribution changes such that $ \mathcal{G}(\tau) = - \int d\omega \frac{e^{-\omega \tau}}{1 - e^{-\frac{\hbar \omega}{k_B T}}}A(\omega)$. }  
As $\mathcal{K}(\tau, \omega)$ vanishes exponentially for large $|\omega|$, 
\begin{equation}
    \lim_{\omega \rightarrow \pm \infty} \mathcal{K}(\omega) \rightarrow 0,
\end{equation}
the inverse of the kernel diverges exponentially in the high frequency limit, rendering the inverse problem exponentially unstable.

In this work, we will construct a convolutional neural network to learn the analytic continuation of imaginary time Green's functions to the real frequency domain. Several other wide-spread approaches to this problem exist so far. Two common approaches are the usage of the MaxEnt algorithm \cite{Jarrell96,Linden95} or stochastic analytical continuation \cite{Beach04a,Sandvik98,Shao23}.  Both approaches are based on a  $\chi^2$-quality  of the analytic continuation as  evaluated by 
\begin{equation}
    \label{eq:chi2}
    \chi^2[A] = \int d\tau d\tau' \left( \int d\omega \mathcal{K}(\tau, \omega) \Hat{A}(\omega) - \mathcal{G}(\tau) \right)C^{-1}(\tau,\tau')
    \left( \int d\omega \mathcal{K}(\tau', \omega) \Hat{A}(\omega) - \mathcal{G}(\tau') \right),
\end{equation}
where $C(\tau,\tau')$ corresponds to the covariance matrix  that accounts  for  correlation in imaginary time in the data.   Within a stochastic evaluation of the imaginary time Green function, the  Green function is given by  ${\cal G}(\tau)   = \sum_{n} P(n) G_n(\tau) \simeq    \frac{1}{N_{\text{Samples}}} \sum_{m=1}^{N_{\text{Samples}}} G_{m}(\tau)  $, where   $n$ denotes an arbitrary configuration space with a probability distribution $P(n)$ and $N_{\text{Samples}}$ corresponds to the number of independent   configurations, $m$,  drawn  from this probability distribution. In this context,  the covariance matrix is 
given by \cite{ALF_v2.4}
\begin{equation}
 C(\tau,\tau')  = \frac{1}{N_{\text{Samples}}} \sum_{m=1}^{N_{\text{Samples}}} \left[ G_m(\tau\phantom{'})  - {\cal G} (\tau) \right]    \left[ G_m(\tau')  - {\cal G} (\tau') \right]
\end{equation} 
In the absence of correlations  in imaginary time,  the covariance 
matrix is diagonal and reduces to the square  of the stochastic error. 


MaxEnt involves a relative entropy functional given by
\begin{equation}
    \mathcal{S}[A] = -\int d\omega \Hat{A}(\omega) \ln{\left(\frac{\Hat{A}(\omega)}{D(\omega)}\right)},
\end{equation}
with $D(\omega)$ the so-called default model encoding any priori knowledge about the spectral density $\Hat{A}(\omega)$\cite{Jarrell96,Linden95}. MaxEnt then minimizes the combination 
\begin{equation}
    Q = \chi^2 -\alpha^{-1} \mathcal{S},
\end{equation}
with $\alpha^{-1}$ being a hyperparameter that is determined by the  algorithm.   In this work, we 
will use the so called classic implementation of  MaxEnt  \cite{Jarrell96,Linden95}. While this procedure allows for good control of the output, 
it has difficulty resolving high energy structures and detailed features of the spectral function. Many  alternatives to the MaxEnt method have been proposed,  including a stochastic approach  \cite{Sandvik98,Beach04a,Shao23}  as well as  methods based on Nevanlinna's theorem \cite{Jiani21}. The very difficulty  of this problem calls  for  alternatives  that can be used for  comparison, 
and further implementations using machine learning algorithms \cite{regression_kernel, linear_kernel}, including neural networks \cite{Implementation1, nn2, nn3, nn4}, can yield some advantages \cite{Implementation1, regression_kernel}, in particular in noise handling.

In general, trained in a supervised manner, neural network approaches are well suited to tackle this kind of ill-posed problems \cite{Koji}.  However, approaches differ in their structure in using dense layers only \cite{nn4}, using convolution layers \cite{nn2, nn3}, or applying Softmax as a method for normalization \cite{Implementation1}. Differently from \cite{nn4}, we use convolutional layers. While \cite{nn2, nn3} also use convolutional layers, in their approach, the  convolution is applied subsequent to the dense layers, breaking the locality and translation invariance of the underlying data.\\

This paper is organized as follows:
In sec. \ref{sec:computational_aspects}, the technical details of our implementation, as well as the generation of training and test data, and the process of training are explained. Subsequently, in sec. \ref{sec:performance}, the results of the trained network are compared to MaxEnt. In sec. \ref{sec:performance:comp2maxent}, we will examine the behavior of the network and its decision making process. In particular, our network is used to detect spin-charge separation in a 1D Hubbard model in sec. \ref{sec:performance:spincharge_sep} \cite{Preuss94,Abendschein06}, and applied 
to the two-dimension Su-Schrieffer-Heeger model on a square lattice \cite{Su80,Goetz21}
 in sec. \ref{sec:SSH}. We end with a conclusion and outlook in sec.~\ref{sec:conlusion}.

%% file: chapters/methods.tex
\section{Computational aspects} 
\label{sec:computational_aspects}
A neural network, i.e. a network of repeating units called neurons, can be seen as a universal function approximator \cite{ML_intro}
\begin{equation}
    f: \mathbb{R}^n \rightarrow \mathbb{R}^m,
\end{equation}
where $n, m \in \mathbb{N}$ describe the dimension of the input and output vector, respectively. The propagation of an input vector $\mathbf{x} \in \mathbb{R}^n$ can be described as
\begin{equation}
    \boldsymbol{a} = \phi(\boldsymbol{W}\cdot \boldsymbol{x} + \boldsymbol{b}),
\end{equation}
where $\boldsymbol{a}$ describes the activation in the layer, $\boldsymbol{x}$ the activation of the previous layer, $\boldsymbol{W}$ the weight matrix, $\phi$ a non-linear activation function, and $\boldsymbol{b}$ the bias.  $\boldsymbol{W}$ and $\boldsymbol{b}$ are the free parameters of the model, which are adjusted while training. 
For training these networks in a supervised manner, a data set $\mathcal{D}$ of $N$ samples, 
\begin{equation}\label{eq:dataset1}
    \mathcal{D} = \{(\boldsymbol{x}_0, \boldsymbol{y}_0), (\boldsymbol{x}_1, \boldsymbol{y}_1), ..., (\boldsymbol{x}_N, \boldsymbol{y}_N)\},
\end{equation}
is used, with  $\boldsymbol{y} \in \mathbb{R}^M$ being the output for their corresponding input vectors $\boldsymbol{x}\in \mathbb{R}^N$.  
The prediction of the network $\boldsymbol{\hat{y}}$ is interpreted as a function of its free parameters such that $\boldsymbol{\hat{y}}(\boldsymbol{w})$, where $\boldsymbol{w}$ collectively denotes the bias $\boldsymbol{b}$ and the weight matrix $\boldsymbol{W}$. 
While training, the problem specific loss $\mathcal{L}(\boldsymbol{y}, \boldsymbol{\hat{y}})$ is minimized using an optimization algorithm on the free parameters $\boldsymbol{w}$. 
Examples of commonly used optimization algorithms are stochastic gradient descent \cite{SGD} or Adam \cite{Adam}, with the previous one being employed in our work.

\subsection{Generating the training data}

In the following we want to train an neural network to invert eqn.~\eqref{eq:ill-posed-problem} which, as stated in the introduction, is an exponentially unstable problem. 
For supervised training of our network, a dataset of the form of eqn.~\eqref{eq:dataset1} and size $N$ is used, 
\begin{equation}
    \mathcal{D} = \{\left(\mathcal{G}_0, A(\omega)_0\right), \left(\mathcal{G}_1, A(\omega)_1\right),..., \left(\mathcal{G}_N, A(\omega)_N\right)\},
\end{equation}
where $\mathcal{G}_i$ and $A(\omega)_i$ denote, respectively, the imaginary time domain Greens function and the spectral density function.

Fortunately, even though the inversion of eqn. \eqref{eq:ill-posed-problem} is ill-posed, eqn. \eqref{eq:ill-posed-problem} itself can be straightforwardly evaluated. 
Following \cite{regression_kernel}, we generate our dataset by randomly generating spectral density functions $A(\omega)_i$ and subsequently applying eqn. \eqref{eq:ill-posed-problem} to calculate $\mathcal{G}_i$. More specifically, we construct the spectral density functions from a set of Gaussians. In \cite{regression_kernel}, those Gaussian distributions were chosen to be uniformly distributed in the region of interest. 
In contrast, we artificially increase the overlap between these Gaussian distributions by introducing collision centers. 
For each sample, we randomly generate $n_\text{col} = 4,5,\dots,8$ collision centers, distributed uniformly within the frequency window $\omega_{\text{c}} \in [-7.5, 7.5]$. 
To each collision center we randomly assign $n_\text{Gauss} = 1,2,\dots,12$ Gaussian distributions. 
These Gaussian distributions with uniform randomly chosen variance $\sigma_\text{Gauss} \in [0.1, 3.0]$ are centered at the collision centers with a variance of  $\sigma_\text{col} = 0.01$. To each Gaussian peak with variance $\sigma_\text{Gauss}$ we assign a height $h$ randomly drawn from a Gaussian distribution with mean 
\begin{equation}
    \Bar{h} = \frac{5}{\sigma_\text{Gauss}^3},
\end{equation}
and variance $\sigma_h = 0.1$. This choice of $\Bar{h}$ is motivated by our experience of how realistic data should look like and should be generated. Moreover, we cut off the height such that $h \in [0, 100]$. Note that this is not the final height of the Gaussian in the spectral density function but, as explained below, the height prior to imposing the sum rule normalization. The final height is later rescaled to statisfy the normalization constraint, i.e. the sum rule eqn.~\eqref{eq:normalization}.
\\
To add non-Gaussian aspects to the training data, we randomly add $n_\text{step} =1,2, 3$ step functions to $20\%$ of the generated data. The position of the step is uniform randomly distributed with $\omega_\text{step} \in [-10, 10]$. Its height is randomly drawn from a uniform distribution with
$h_\text{step} \in [-5, 5]$. 
Finally, to reduce the sharpness of the step functions, we low-pass filter the training data with an Infinite Impulse Response (IIR) filter 
\begin{equation}
    \Hat{\omega}(\tau_i) = (1-\gamma)\omega(\tau_i) + \gamma\Hat{\omega}(\tau_{i-1})
\end{equation}
where we randomly start with the smallest or the largest $\tau$. 
The final height is defined by rescaling the entire  spectral density function to ensure the sum rule
\begin{equation}
    \label{eq:normalization}
    \int_{-\infty}^{\infty} A(\omega) d\omega = 1.\\
\end{equation}

As the original distribution of the physical spectral density functions is unknown, we utilize this multi-step process such that the created dataset includes as many semi-realistic spectral density functions as possible. We create $N_\text{train} = 35008$ training and $N_\text{test} = 3200 $ test data for inverse Monte Carlo sampling temperatures $\beta = \frac{1}{k_B T} \in \{10, 20\}$ each. For the final training, we add spatial correlated, i.e. non-white, noise \cite{e25121593} to the data to better mimic the noise output of the Monte-Carlo simulations. For this noise, we randomly select an inverse power-law with exponent $\alpha \in [1, 5]$. 

\subsection{Description of the training setup}
\label{sec:structure}

We now turn to the description of the structure of our neural network, as well as the training procedure used.

\subsubsection{Structure of the neural network}

We use a Convolutional Neural Network (CNN) for predicting the spectral density function. CNNs usually consist of one or multiple convolutional layers \cite{Conv} followed by a dense part. In the context of image recognition, convolutional layers are used as feature finding layers. However, in a more physics driven approach, these layers apply local operators where the locality is controlled by the size of the kernel, and enforce a translation invariance of the trained features. 

\begin{figure}
    \centering
    \hspace*{0mm}(a) \hspace*{51mm} (b) \hspace*{51mm} (c)\\
    \includegraphics[width=\linewidth]{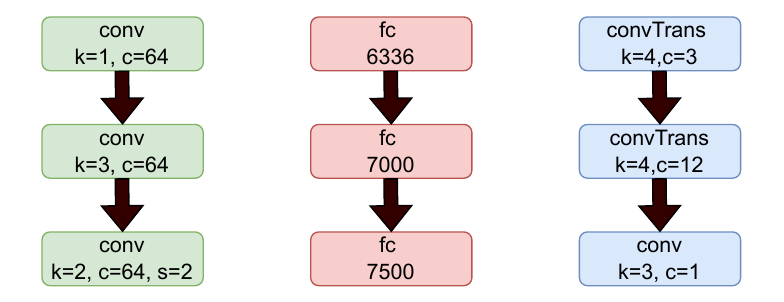}
    \caption{Scheme of the layers in the convolution (a), the dense (b), and deconvolution (c) parts of the neural networks. In (a), the top row denotes the layer type, the first value $k$ describes the kernel size, the second value $c$ the output channel size for this layer, and the optional third value $s$ the stride if different to one. A similar notation applies to (c), where convolutional transposed layers are utilized. For the dense part described in (b) the corresponding layer sizes are shown.}
    \label{fig:network}
\end{figure}

As the input data are artifically made noisy and we want to invert an integral operator, the slopes of the input data, i.e. the first derivatives, are important features in the input Greens functions $\mathcal{G}$. 
As the operation of determining a slope is local and translation invariant we start our network with a set of three convolution layers calculating the slopes. As those operations do not require a bias, we only allow for adjusting the convolution weights while enforcing a zero bias. Moreover, we chose as activation functions for those layers
\begin{equation}
    \textsc{PReLU}(x) = \begin{cases}
x &\text{if } x \geq 0\\
\gamma \cdot x &\text{else}
\end{cases},
\end{equation}
with $\gamma\in\mathbb{R}$ being a trainable parameter.
The convolution part of the neural network can be seen in Fig.~\ref{fig:network} (a).\\

For our neural network simulations, the size of the input data, i.e. the dimensionality of the input data vector is chosen to be $n=101$. The dense part of the CNN consists of three layers with $6336$ nodes in the input, $7000$ nodes in the hidden, and $7500$ nodes in the output layer of the dense part. As activation function, we utilize Rectified Linear Unit (\textsc{ReLU}) activation functions \cite{LReLU}, defined as
\begin{equation}
    \textsc{ReLU}(x) = \begin{cases}
x &\text{if } x \geq 0\\
0 &\text{else}
\end{cases},
\end{equation}
The dense part of the network is depicted schematically in Fig.~\ref{fig:network} (b).\\

Next, we apply a deconvolution, increasing the spatial size to $20,000$ output neurons. Here, we utilize a set of two convolution transposed \cite{5539957} layers as well as a final convolution layer to reduce memory requirements. As before in the convolution, we suppress the bias $\boldsymbol{b}$ and only adjust the weights $\boldsymbol{W}$. However, similar to the dense part of the neural network, we use \textsc{ReLU} activation functions in the deconvolution part of the network. The deconvolution part of the network is depicted in Fig.~\ref{fig:network} (c).\\

Finally, we apply a rescaling layer ensuring the normalization (sum rule \eqref{eq:normalization}) of the output. To this end, the layer calculates the absolute area under the given spectral density function and rescales the result accordingly.

\subsubsection{Loss function and trainer}

To train our network, we utilize a two-fold loss function consisting of a sum of squared errors and a sum of absolute errors\footnote{Both contributions are similar to mean squared error and mean absolut error, respectively, up to a rescaling by the batch size $N$.}. The sum of squared errors is given by
\begin{equation}
\label{eq:mse_close}
    \mathcal{L}_{\text{l2}} = \sum_{i=0}^{N} \sum_{\omega}\left(A_{i}(\omega) - \Hat{A}_{i}(\omega)\right)^2,
\end{equation}
with $N$ being the batch size, i.e. the number of training data in a single batch.  
The sum of absolute errors part is given by
\begin{equation}
\label{eq:mae_close}
    \mathcal{L}_{\text{l1}} = \sum_{i=0}^{N} \sum_{\omega}\left|A_{i}(\omega) - \Hat{A}_{i}(\omega)\right|.
\end{equation}
For large deviations $\left(A_{i}(\omega) - \Hat{A}_{i}(\omega)\right)$ the loss is dominated by the squared error, while for smaller deviations, the loss is dominated by the absolute error. We utilize this combination as the training data consists of peaks with varying height. While larger peaks, in general, are more important to be precise, the error should not be dominated by those larger peaks, causing the network to focus mainly on those large peaks. The final loss function is given as
\begin{equation}
\label{eq:loss_function_total}
    \mathcal{L} = \mathcal{L}_{\text{l2}} + \mathcal{L}_{\text{l1}}.\\
\end{equation}

Finally, we train the neural network for $2,000$ epochs with stochastic gradient descend (SGD) as optimizer, with learning rates decreasing from $10^{-2}$ to $10^{-5}$ until the networks loss saturates.

%% file: chapters/performance.tex
\section{Performance \& physical results} 
\label{sec:performance}
In the following section, the results of our neural network are compared to MaxEnt. In a first step we compare the performance of both approaches on an artificially generated validation set. Subsequently, we apply the network to two physical systems, the one-dimensional Hubbard model and the two-dimensional Su-Schrieffer-Heger (SSH) model,  and discuss the performance of both approaches in recognizing the physically relevant features in those systems.

\subsection{Comparison to MaxEnt on validation data}
\label{sec:performance:comp2maxent}
As a first test of our network, we compare its performance against MaxEnt on validation data. For this purpose, we generate additional $64$ samples of data, which were neither used to train nor to test the network's performance while development. For this data set, the Greens function is calculated from an existing spectral density function, and the predictions of MaxEnt and the network are compared to the known ground truth, i.e. the analytically known result in the real frequency domain. As metrics we apply the mean square error (MSE), the mean absolut error (MAE), and the Wasserstein distance \cite{PhysRevResearch.3.043093}. The MSE is defined as
\begin{equation}
    \textsc{MSE} = \frac{1}{N} \mathcal{L}_\text{l2}, 
\end{equation}
where $N = 64$ and $\mathcal{L}_\text{l2}$ is given in eqn.~\ref{eq:mse_close}. 
Similarly, the mean absolute error (MAE) is given by
\begin{equation}
    \textsc{MAE} = \frac{1}{N} \mathcal{L}_\text{l1},
\end{equation}
where $\mathcal{L}_\text{l1}$ is given in eqn.~\ref{eq:mse_close}. Note that our neural network was trained to minimize MSE and MAE as they are given (up to scaling) by the respective loss functions \eqref{eq:mse_close} and \eqref{eq:mae_close}.

Finally, in order to have another independent distance measure not built into the loss function, we employ the Wasserstein distance \cite{PhysRevResearch.3.043093} as an independent metric, defined by
\begin{equation}
    W_1(A, \Hat{A}) = \inf_{\pi \in \Gamma(A, \Hat{A})} \int_{\mathbb{R} \times \mathbb{R}} |\omega - \Hat{\omega}| \, d\pi(\omega, \Hat{\omega}),
\end{equation}
where $\Gamma(A, \Hat{A})$ is the set of all couplings (joint distributions) with marginals $A$ and $\Hat{A}$. Moreover, we write for $A_{i}(\omega),\, \Hat{A}_{i}(\omega)$ the short-hand notation $A,\, \Hat{A}$, respectively\footnote{Also see \url{https://docs.scipy.org/doc/scipy/reference/generated/scipy.stats.wasserstein_distance.html}.}. Naively speaking, the Wasserstein distance quantifies the minimal transport cost, given by how large $|\omega - \Hat{\omega}|$ needs to be to transform $\Hat{A}$ into $A$.\\

The results of this test are shown in Table~\ref{tab:comparision}. As can be seen, the network and MaxEnt perform roughly similar when reconstructing the spectral density function, with a slight but consistent advantage for the network on all metrics. Please note that the spectral density functions used for this comparision are generated similarily to the training and test data of the neural network and thus perfectly fit into the distribution of functions the network is trained for.\\

\begin{table}[]
    \centering
    \begin{tabular}{c||c|c}
        Metric & Network & MaxEnt \\
        \hline \hline
        MSE\,$\downarrow$ & 0.108 & 0.121 \\
        \hline
        MAE\,$\downarrow$ & 0.896 & 0.950 \\
        \hline
        Wasserstein distance\,$\downarrow$ & 0.511 & 0.559
    \end{tabular}
    \caption{Comparison of predictions by the neural network and MaxEnt to a ground truth based on a set of 64 randomly generated spectral density functions. While both algorithms perform qualitatively similar, the network consistently has a slight advantage over all metrics when applied to generated spectral functions similar to the training data. The arrow $\downarrow$ next to the name of each metric indicates that smaller values mean better performance.}
    \label{tab:comparision}
\end{table}

{In Fig.~\ref{fig:examples}, three exemplary predictions of our artificially generated data are shown.
For each example, the expected spectral density function labeled 'True', as well as predictions by MaxEnt and the neural network, are provided.
Here, the systematic differences between the predicitons by MaxEnt and the neural network become evident.
In general, our network seems to be better than MaxEnt at predicting the exact location of a high peak as can be seen in Fig.~\ref{fig:examples_a} and Fig.~\ref{fig:examples_b} on the right peak, as well as in Fig.~\ref{fig:examples_c}.
However, MaxEnt proves to be more sensitive to small peaks, which our network sometimes ignores, c.f.~e.g. in Fig.~\ref{fig:examples_a} the shallow 'hill' next to the highest peak and the small but steep peak in the left of Fig.~\ref{fig:examples_b}.
The reason for this bias is the loss function described in Eq.~\ref{eq:loss_function_total}. 
As the loss function compares the difference in height of the predicted and actual spectral density function for each $\omega$, the loss prioritizes the exact location over the exact height, especially for sharp peaks. 
This raises the question whether a loss function considering spatial relations in $\omega$ such as the Wasserstein distance might be better suited for the training of the neural network.
Another noteworthy aspect is the different pattern of overfitting in our neural network, as compared to  MaxEnt. 
While the network tends to ignore smaller peaks, MaxEnt overfits aspects of the images as can be seen in Fig.~\ref{fig:examples_b} in the smaller peak on the left, where  MaxEnt also produces oscillations,  and similarly for the large peak on the left in Fig.~\ref{fig:examples_c}.



\begin{figure}[t]
    \centering
    \begin{subfigure}[t]{0.60\textwidth}
        \centering
        \includegraphics[width=\linewidth]{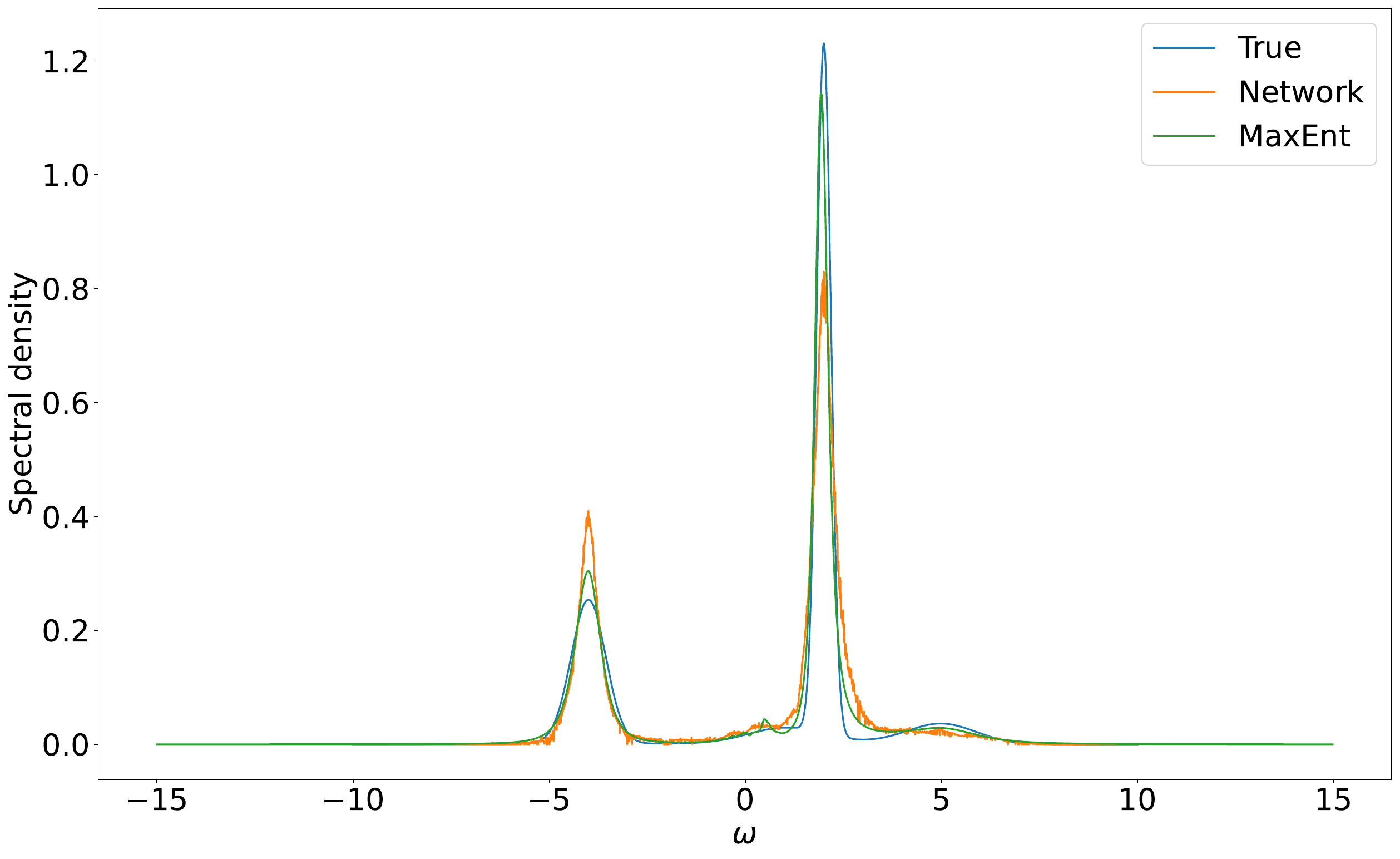}
        \caption{}
        \label{fig:examples_a}
    \end{subfigure}
    \begin{subfigure}[t]{0.60\textwidth}
        \centering
        \includegraphics[width=\linewidth]{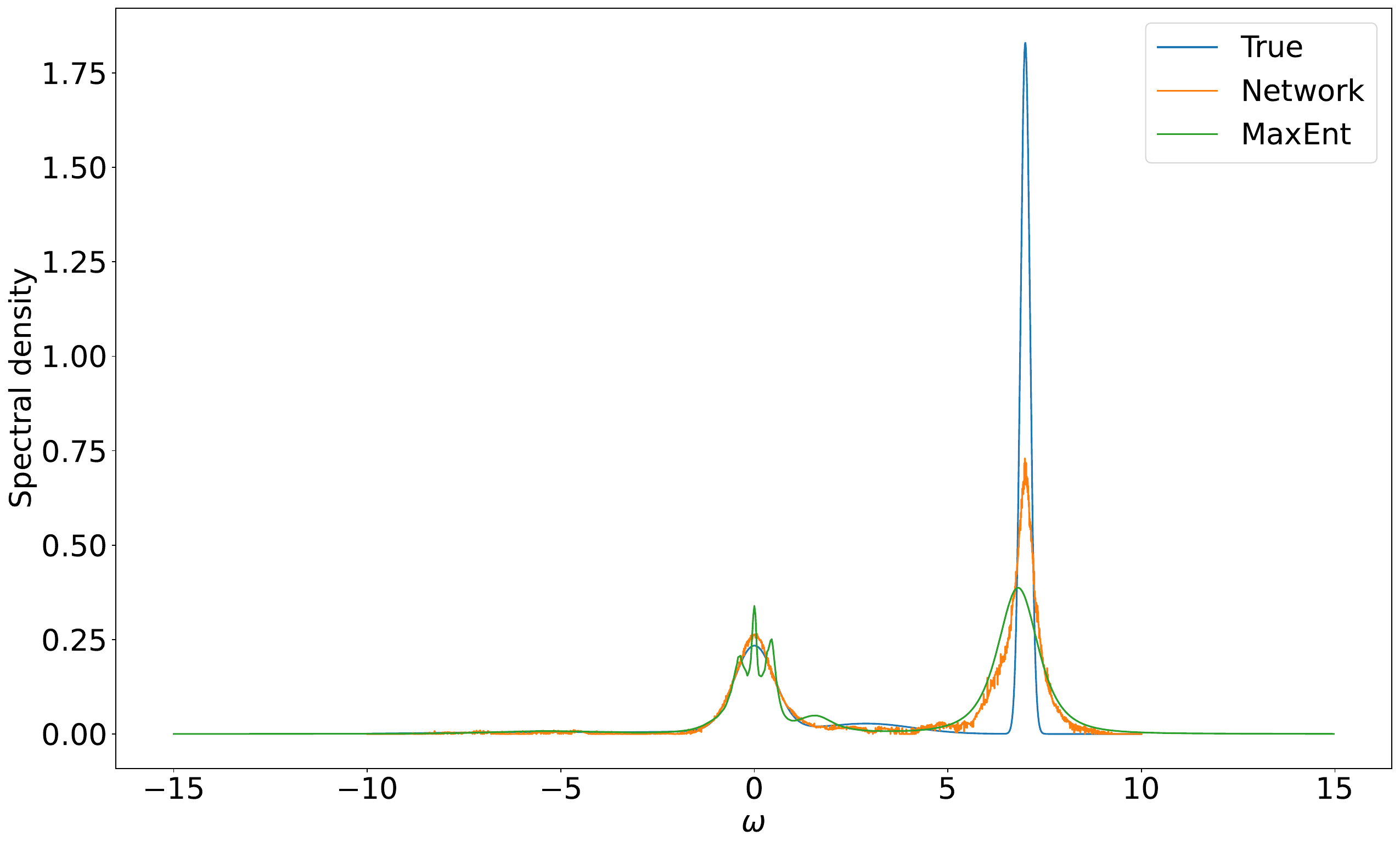}
        \caption{}
        \label{fig:examples_b}
    \end{subfigure}
    \begin{subfigure}[t]{0.60\textwidth}
        \centering
        \includegraphics[width=\linewidth]{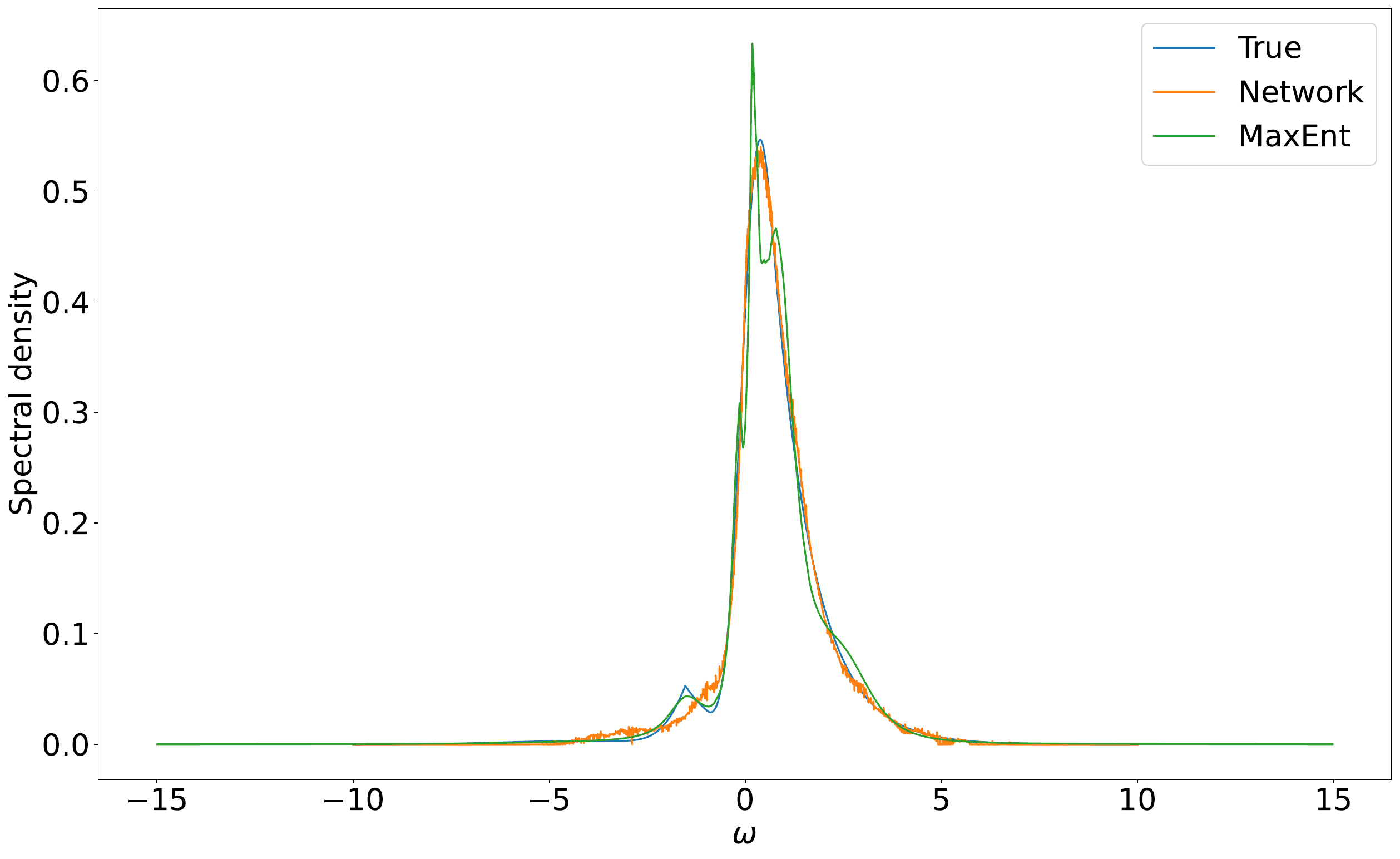}
        \caption{}
        \label{fig:examples_c}
    \end{subfigure}
    \caption{{Exemplary predictions for artificially generated data.  Each example includes the true (generated) spectral density function as well as predictions by MaxEnt and the neural network. 
    Systematic differences between the predictions of MaxEnt and the neural network can be seen in the prediction of heights and positions of peaks, as well as MaxEnt inducing oscillations around sharp peaks.}
    }
    \label{fig:examples}
\end{figure}

\clearpage
\subsection{Spin-Charge separation in 1D Hubbard model}
\label{sec:performance:spincharge_sep}
In one spatial dimension, the  Fermi liquid is unstable to interactions. The electron  with  spin 1/2  and  unit electric charge  fractionalizes  
into two entities, the  chargeless spinon  carrying the spin 1/2 degree of freedom, and the spinless holon carrying the charge \cite{Giamarchi}.  Importantly both entities have different velocities, and the spectral function of the electron is a convolution of the spinon and holon spectral functions.  This fractionalization of the physical electron  is best seen in 
dynamical correlation functions, for which analytical continuation is often feasible. Beyond the well-established phenomenon of spin-charge separation in one-dimensional (1D) Mott insulators, first understood via the exact Bethe ansatz solution of the Hubbard model \cite{LiebWu1968}, recent theoretical and experimental work has revealed the possibility of spin-orbital separation, wherein an injected electron can decay into three independently propagating quasiparticles: a spinon (carrying spin), an orbiton (carrying orbital excitation), and a holon (carrying charge). Using a microscopic model for the quasi-1D compound Sr$_2$CuO$_3$, \cite{Ritschel2006} demonstrated theoretically that spin and orbital degrees of freedom dynamically decouple. This prediction was subsequently confirmed experimentally in \cite{Schlappa2012} using resonant inelastic x-ray scattering, revealing clear evidence for the independent propagation of spinons and orbitons.

The Hubbard model that we consider here  does not take into account orbital degrees of freedom. Its Hamiltonian reads
\begin{equation}
\hat{H}  =  -t \sum_{i,\sigma} \left( \hat{c}^{\dagger}_{i,\sigma} \hat{c}^{\phantom\dagger}_{i+1,\sigma} + \hat{c}^{\dagger}_{i+1,\sigma} \hat{c}^{\phantom\dagger}_{i,\sigma}\right) + \frac{U}{2}  \sum_i \left( \hat{n}_i -1 \right)^2.
\end{equation}
Here, $i$  labels the sites of a one-dimensional chain, $\hat{c}^{\dagger}_{i,\sigma}$ creates an electron  with z-component of spin $\sigma= \uparrow,\downarrow$ in 
a Wannier state centered around site $i$,  and $\hat{n}_i  =  \sum_{\sigma} \hat{c}^{\dagger}_{i,\sigma} \hat{c}^{\phantom\dagger}_{i,\sigma}$ is the 
number operator. We carry out calculations  on an $L=46$  site chain, at  $U/t=4$  and $\beta t  = 10$, with $\beta$ being the inverse temperature.   At vanishing chemical potential, particle-hole symmetry
ensures half-filling, $\frac{1}{L}\sum_{i=1}^{L} \langle \hat{n}_i \rangle  = 1$.  We have carried out the calculations with the Algorithms for Lattice fermions (ALF) package \cite{ALF_v2.4}  for the finite temperature auxiliary field quantum Monte Carlo algorithm \cite{Blankenbecler81,White89}.

The key feature of the data presented in Fig.~\ref{fig:performance_physical_results:comp_2_MaxEnt:spin-charge}(b) is a gap at wave vector $k=\pi/2$.  As mentioned above, a  doped hole at negative energies splits into a spinon and a holon.  In the simplest possible understanding,  the spectral function of the electron is  given by  a convolution of that of the spinon and of the holon. As such, the spectral function   is characterized by  the absence of poles and structure   distribution 
of the spectral weight.  In particular, at  $k< \pi/2$   we observe  two branch cuts corresponding
to the fractionalized excitations.  It is challenging to resolve them as a function of $k$. At $k=0$ we clearly observe two features at 
$\omega \simeq -1.6$ and $\omega \simeq -3.0$  corresponding to the spinon and holon, respectively.  Both excitations are expected to merge 
together at $k = \pi/2$. Other features in the spectral function have been discussed intensively in the literature and it is beyond the scope of this article to account for the details of the physics. We refer the interested reader to Ref.~\cite{Abendschein06} for more information.



\begin{figure}[!ht]
    \centering
    \begin{subfigure}[t]{0.60\textwidth}
        \centering
        \includegraphics[width=\linewidth]{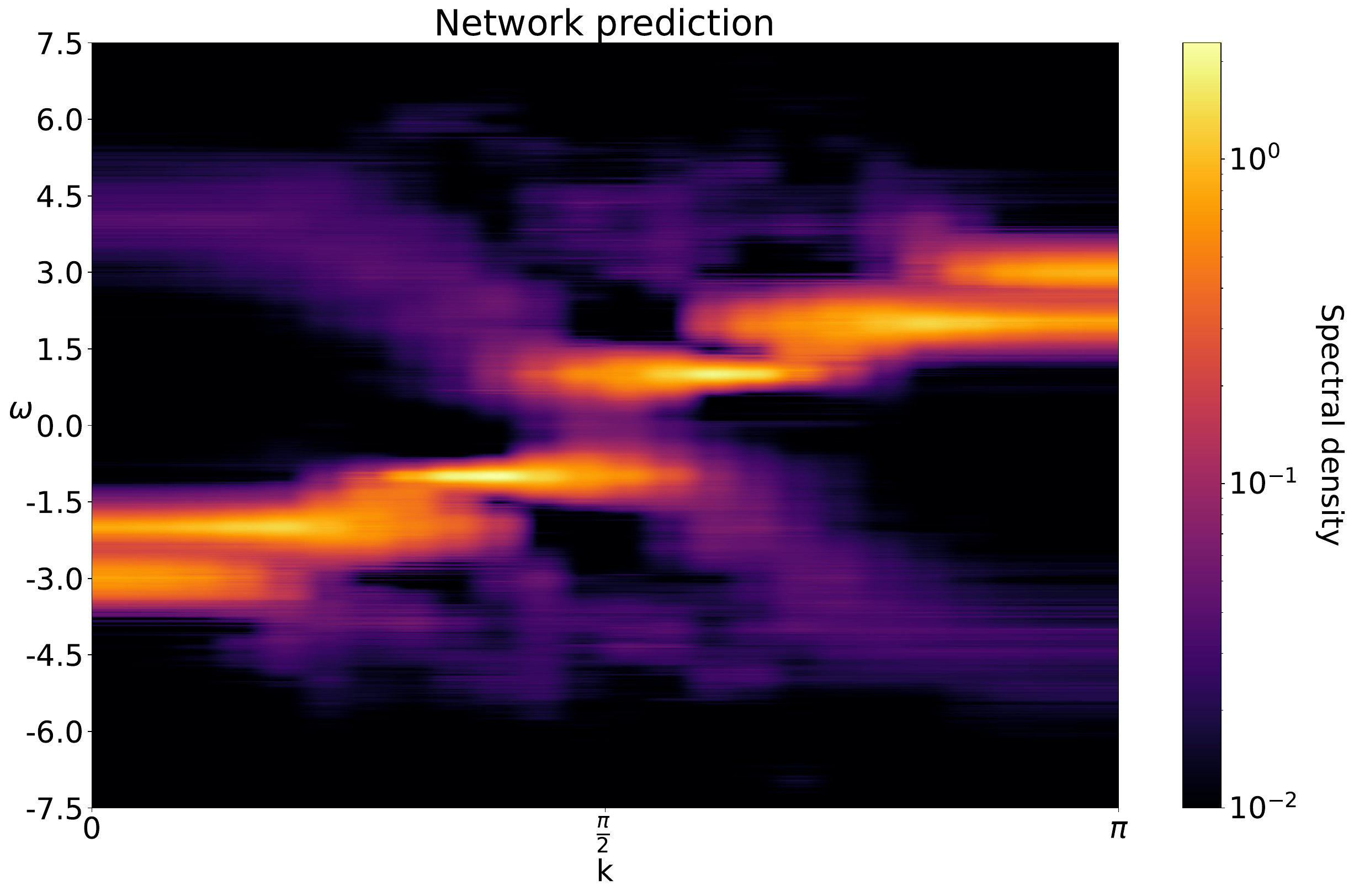}
        \caption{Prediction of the spin--charge separation by the network}
        \label{fig:performance_physical_results:comp_2_MaxEnt:spin-charge_a}
    \end{subfigure}
    \begin{subfigure}[t]{0.60\textwidth}
        \centering
        \includegraphics[width=\linewidth]{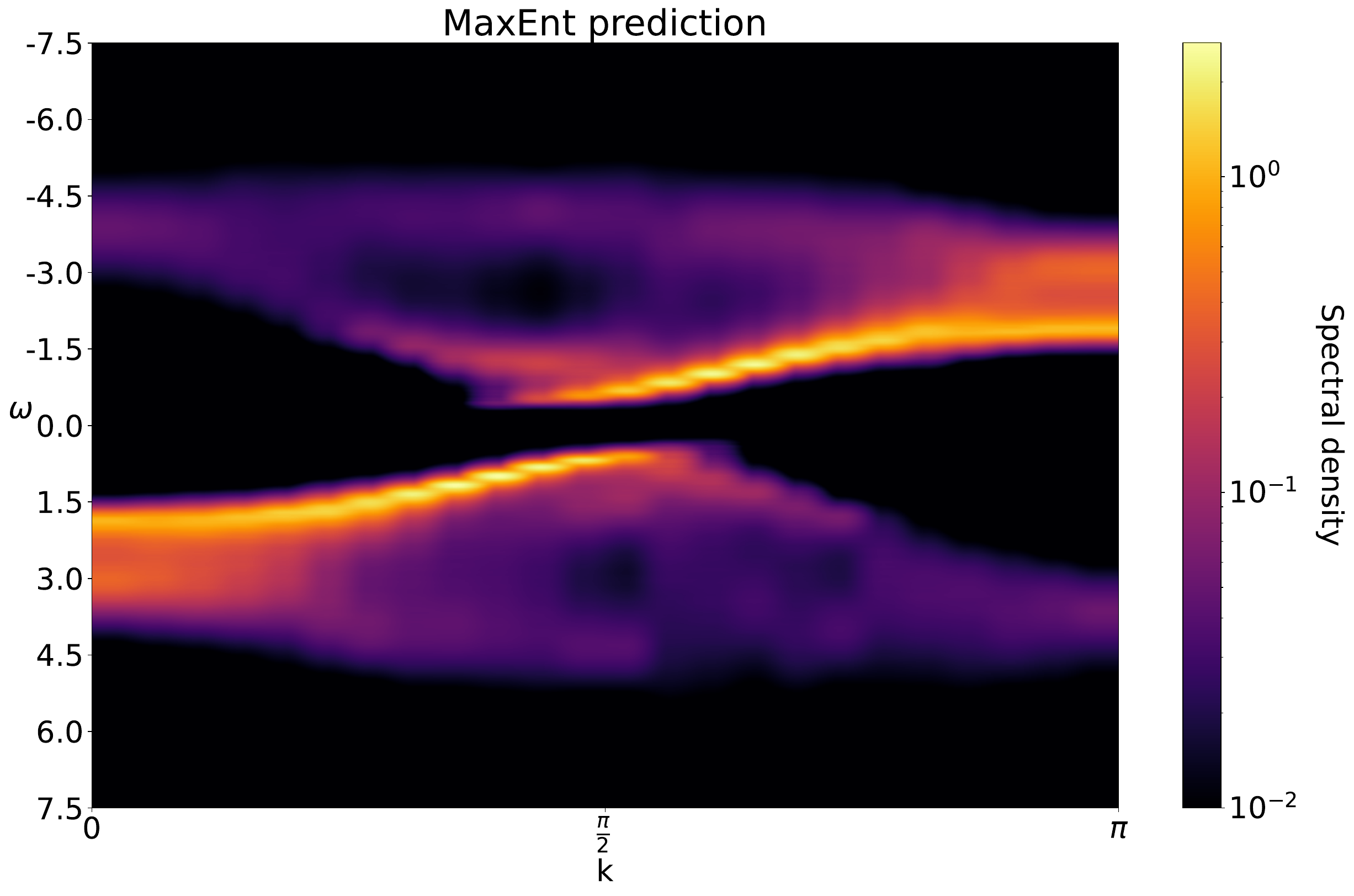}
        \caption{Prediction of the spin--charge separation by MaxEnt}
        \label{fig:performance_physical_results:comp_2_MaxEnt:spin-charge_b}
    \end{subfigure}
    \caption{Predicted spectral density functions for the 1D Hubbard model using the neural network \ref{fig:performance_physical_results:comp_2_MaxEnt:spin-charge_a} and MaxEnt \ref{fig:performance_physical_results:comp_2_MaxEnt:spin-charge_b}. Here we consider an $L=46$ site chain  at $U/t=4$  and $\frac{t}{k_bT} = 10$ at half filling.  The MaxEnt plot is taken from \cite{ALF_v2.4}. While in both images spin-charge separation is discernible, the network  produces spurious {stratification} of the image. 
    \label{fig:performance_physical_results:comp_2_MaxEnt:spin-charge}}
\end{figure}

In this work, we use data for the spectral function of the 1D Hubbard model  obtained using the ALF package \cite{ALF_v2.4} 
to benchmark our neural network analytic continuation against the established MaxEnt method.  
Fig.~\ref{fig:performance_physical_results:comp_2_MaxEnt:spin-charge}(a)  plots the neural network analytic continuation results.   In comparison to  MaxEnt we observe   a loss of spectral  weight at positive (negative)  frequencies   and negative  (positive) momenta,   corresponding  to the so called shadow bands
\cite{Penc96}. 
The Hubbard model at half-filling has dominant anti-ferromagnetic correlations  such that  a mode  at a wave vector $k$ can scatter off a gapless spin-excitation carrying momentum $q=\pi$,  and hence  generate a shadow at momentum  $k + \pi$.  
Finally, we observe {stratification} of the image produced by the network, which introduces additional spurious features. 

\subsection{The two-dimensional SSH model: self-consistent Born approximation}
\label{sec:SSH}

\begin{figure}
    \centering
    \begin{subfigure}[t]{0.60\textwidth}
        \centering
        \includegraphics[width=\linewidth]{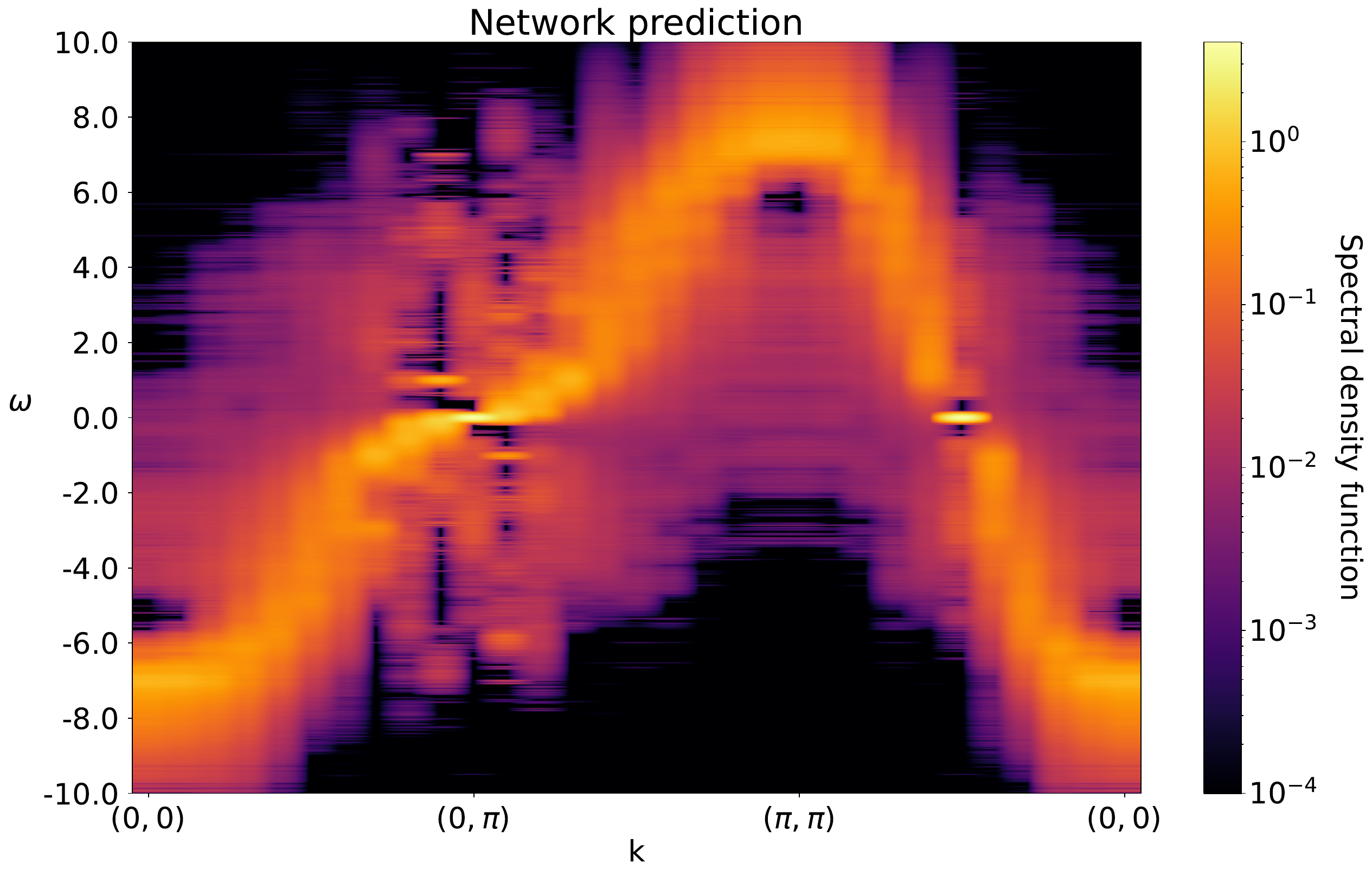}
        \caption{Prediction for the 2D SSH model by the network.}
        \label{fig:SSH_a}
    \end{subfigure}
    \begin{subfigure}[t]{0.60\textwidth}
        \centering
        \includegraphics[width=\linewidth]{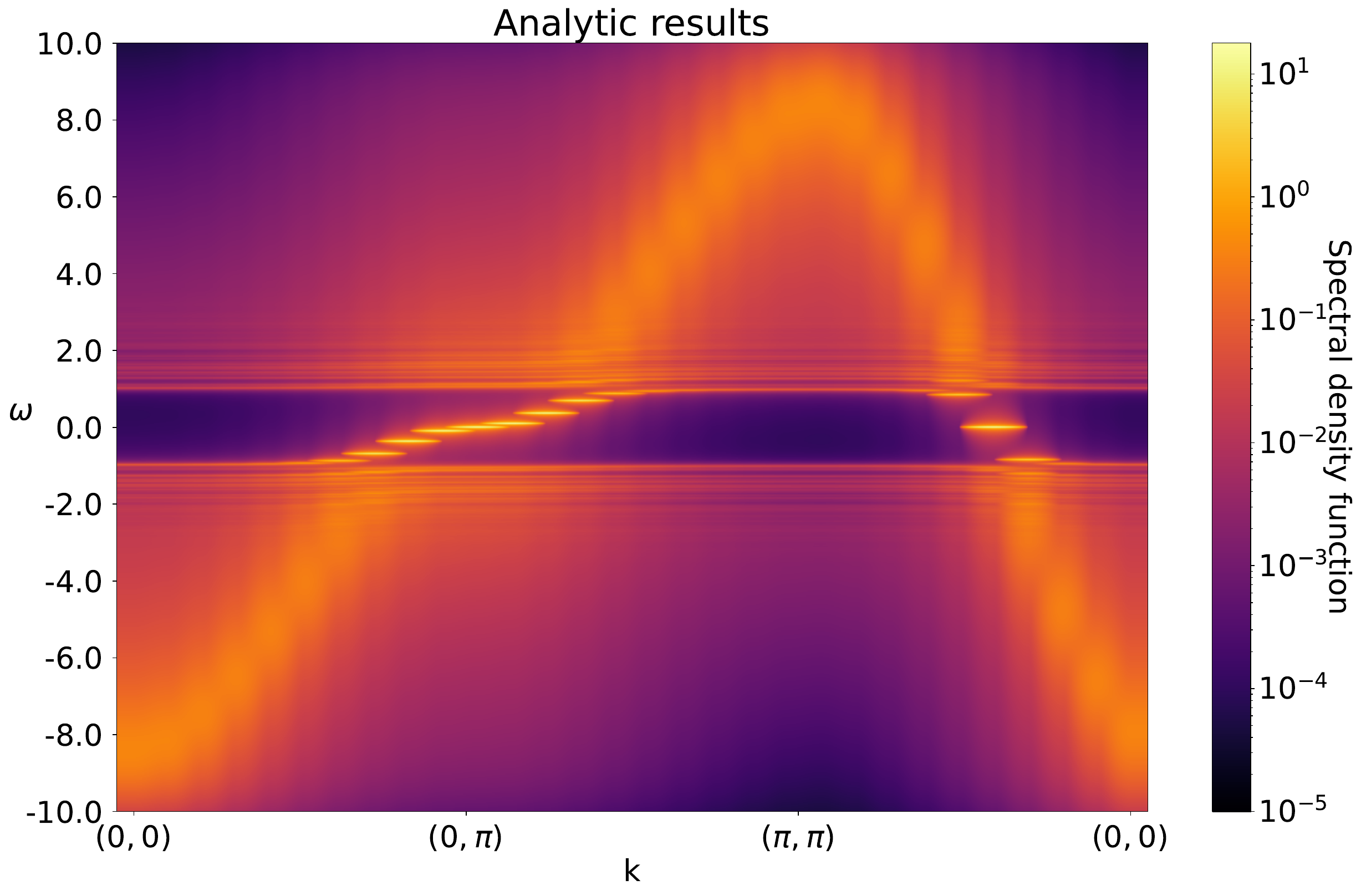}
        \caption{Analytic results for the 2D SSH model.}
        \label{fig:SSH_b}
    \end{subfigure}
    \caption{ Fig. \ref{fig:SSH_a} shows the neural network prediction for the spectral density of the 2D SSH model. 
    Fig. \ref{fig:SSH_b} is reproduced from Fig. 12(a) of  \cite{Goetz21}. The data  presented here \cite{Goetz21} corresponds  to the half-filling case  at $\omega_0 = \sqrt{\frac{k}{m}} = 1.0$,  $k_b T = 1/40$, $t = 1.85$  and $g = 1.5$. $\ve{k}$ runs over a path in the Brillouin zone,  from  $(0,0)$ to $(0,\pi)$ to $(\pi,\pi)$  and back to $(0,0)$.  The  frequency ranges from $-10$  to  $10$  in steps of $0.01$.}
    \label{fig:SSH}
\end{figure}

As  a  second benchmark  case,  we consider a specific version of the  two-dimensional Su-Schrieffer-Heeger (SSH) model \cite{Su80},  given by
\begin{equation}
\hat{H} =   \sum_{ b = \langle \ve{i}, \ve{j} \rangle } \left(-t  + g \hat{X}_b\right)  \left( \sum_{\sigma} \hat{c}^{\dagger}_{\ve{i},\sigma} 
\hat{c}^{\phantom\dagger}_{\ve{j},\sigma} + \hat{c}^{\dagger}_{\ve{j},\sigma} 
\hat{c}^{\phantom\dagger}_{\ve{i},\sigma}\right)   +  \sum_{b}  \frac{\hat{P}^2_b}{2m} + \frac{k}{2}\hat{X}^2_b.
\end{equation}
On each nearest neighbor bond $b$ of a square lattice with sites  labelled by the $\ve{i}$, $\ve{j}$ sits a harmonic oscillator  that dynamically  
modulates the hopping matrix element.   While it is possible to carry out Monte Carlo simulations of this model \cite{Goetz21,Goetz23}, we  adopt a self-consistent Born approximation described in Ref.~\cite{Assaad08,Goetz21}.   Importantly, this approximation goes a good way at capturing some salient features 
of  the single particle spectral function.  It also allows for a calculation on the real time axis,  such that no analytical continuation is  required.  
For  comparison  with the neural network analytical continuation  we transform the real  frequency data  to imaginary time -- this transformation is numerically stable -- and then  use  the neural network to transform back to   real frequencies.

The data  presented in Fig.~\ref{fig:SSH}b   is  taken  from \cite{Goetz21} and corresponds  to the half-filling case  at $\omega_0 = \sqrt{\frac{k}{m}} = 1.0$,  $k_b T = 1/40$, $t = 1.85$  and $g = 1.5$.  
As apparent, below the phonon frequency, $|\omega| < \omega_0 $, we observe a well-defined renormalized quasi-particle excitation. 
Above  $\omega_0$, the coupling to the phonon degrees of freedom provide a substantial broadening of  the bare dispersion relation, resulting in a substantial incoherent background.
The  data provides an interesting test case  where coherent and incoherent excitations are to be resolved simultaneously.  {Upon comparison,  we observe that the neural  
network can  resolve the incoherent high energy features of the image as well as, to some degree, the low energy coherent quasi-particle excitations inside the gap, but fails to properly resolve the infrared gap itself. This is, however, not suprising, as our network is trained with data built from Gaussians, and not with data that explicitly included gaps, the latter of which hence are out of distribution samples.}

%% file: chapters/discussion.tex
\section{Conclusion and Outlook}\label{sec:conlusion}

{In this work, we employed a neural network to predict a spectral density function from a given imaginary-time Green’s function.
Training a network for this task poses several challenges, primarily due to the ill-posed nature of analytical continuation and the presence of noise in imaginary-time Green’s functions. One of the central obstacles for applying machine learning to such problems is the construction of suitable training data.
Due to the large variability of possible spectral density functions, we developed a multi-stage procedure for the automatic generation of artificial spectral densities.

In more detail, we proposed methods for generating high-quality training and testing data with sufficient complexity. The training data predominantly consist of Gaussian peaks. To increase the probability of peak overlap, these peaks were generated around predefined collision centers, with peak positions drawn from a normal distribution centered at each collision point. To incorporate non-smooth and unsteady features, we randomly added Heaviside step functions with both positive and negative amplitudes. In addition, a subset of the data was randomly low-pass filtered to introduce further complexity and heterogeneity. This procedure yields a rich and diverse set of spectral density functions for training the network.
Furthermore, we added artificial pink noise to the data in order to closely mimic the correlated noise arising from Monte Carlo simulations of Green’s functions. 

We evaluated the performance of the neural network by comparing it to MaxEnt, as well as by applying it to established physical models, including spin--charge separation in the one-dimensional Hubbard model, as well as the  SSH model on a  square lattice  within the self-consistent Born approximation.
Using an additional set of artificially generated validation data, we found that the chosen loss function enforces a stricter localization of peak maxima, while being less accurate in predicting peak heights than MaxEnt.
Nevertheless, on this validation data, the network outperforms MaxEnt across all tested metrics, including the mean squared error and the Wasserstein distance. This highlights the potential of neural network based approaches for addressing ill-posed inverse problems.
For physically motivated tests, most notably spin--charge separation in the one-dimensional Hubbard model studied using auxiliary-field quantum Monte Carlo methods \cite{ALF_v2.4}, as well as the electron--phonon problem treated within the self-consistent Born approximation \cite{Goetz21}, the network currently performs worse than MaxEnt.

We therefore conclude that, while the artificial training data partially capture relevant physical features, insufficient coverage remains a limiting factor for network performance. This discrepancy indicates a systematic difference between artificial and real data that has yet to be resolved within the data generation process.
Despite these limitations, the results suggest that further improvements in the training data generation scheme could enhance the network’s capabilities beyond those of MaxEnt. At present, the dominant limitation arises from the data generation procedure and the presence of out-of-distribution samples when applying the network to physical data.

 In practice, the network provides an easy-to-use tool that already yields reasonably accurate identification of physically relevant features. More broadly, our results demonstrate that neural networks are, in principle, capable of solving this task and point toward future research directions with the goal to further improve the network's capabilities. Beyond artificial data generation, the network could also be trained on experimental spectroscopic data available online and in established databases. Real-frequency spectral data, once digitized, can be transformed into imaginary-time data in the stable direction of the problem. The resulting paired data sets of imaginary- and real-frequency information can then be used directly for training.
Experimental data may further be complemented by results from various approximation methods such as the self-consistent Born approximation discussed in this work to further enrich the training set. In summary, this raises an important  question of how much data are required to train a neural network that consistently outperforms standard analytical continuation methods?
}